\documentclass[12pt]{article}
\usepackage{epsfig}
\usepackage{cite}

\newcommand{\mysection}{\setcounter{equation}{0}\section}

\def\beq{\begin{equation}}
\def\eeq{\end{equation}}
\def\beqa{\begin{eqnarray}}
\def\eeqa{\end{eqnarray}}
 
\newlength{\dinwidth} \newlength{\dinmargin}
\begin{document}

\begin{center}
{\Large \bf Higher-order corrections for $tZ$ production via anomalous couplings}
\end{center}
\vspace{2mm}
\begin{center}
{\large Nikolaos Kidonakis}\\
\vspace{2mm}
{\it Department of Physics, Kennesaw State University,\\
Kennesaw, GA 30144, USA}
\end{center}
 
\begin{abstract}
I present higher-order corrections from soft-gluon 
emission for the associated production of a top quark with a $Z$ 
boson via anomalous $t$-$q$-$Z$ couplings at LHC energies. 
Approximate next-to-next-to-leading order (aNNLO) double-differential cross sections are derived from soft-gluon resummation at next-to-leading-logarithm accuracy. The total $tZ$ production cross sections and the top-quark transverse-momentum and rapidity distributions are calculated at aNNLO for various LHC energies. It is shown that the soft-gluon corrections are dominant and large for this process. 
\end{abstract}
 
\mysection{Introduction}
 
The study of the top quark is a very important part of physics at the Large Hadron Collider (LHC). As the heaviest known elementary particle, the top quark has unique properties and is central to the elucidation of the Higgs mechanism. Many reviews of top quark physics have been written, see for example Refs. \cite{WB,DW,IQWW,AHTJ,NKBP,RBAF,PF,NKHQ13,DDL,AG,JWK,AGRS}, that describe top-quark production and top-quark properties. The main production process is top-antitop pair production \cite{NKtop} followed by single-top production \cite{NKts,NKtW}.

In addition to the various Standard Model processes for top production, another possibility is production via top-quark anomalous couplings. One such process is the associated production of a top quark with a $Z$ boson, see e.g. \cite{TY,NKAB,LZ,AAC,TopC,DMZ}. While $tZ$ production may proceed in the Standard Model via processes involving an additional quark in the final state \cite{SMtZ}, in models of new physics with anomalous top-quark couplings it is possible to produce a $tZ$ final state without any additional particles at lowest order.

An effective Lagrangian that includes the coupling of a $t,q$ pair to a $Z$ boson in such processes may be written as 
\begin{equation}
\Delta {\cal L}^{eff} =    \frac{1}{ \Lambda } \,
\kappa_{tqZ} \, e \, \bar t \, \sigma_{\mu\nu} \, q \, F^{\mu\nu}_Z + h.c.,
\label{Langrangian}
\end{equation}
where $\kappa_{tqZ}$ is the anomalous $t$-$q$-$Z$ coupling, with 
$q$ a $u$- or $c$-quark;
$F^{\mu\nu}_Z$  is the $Z$-boson field tensor;
$\sigma_{\mu \nu}=(i/2)(\gamma_{\mu}\gamma_{\nu}
-\gamma_{\nu}\gamma_{\mu})$ with $\gamma_{\mu}$ the Dirac matrices;
$e$ is the electron charge; and $\Lambda$ is an effective scale which 
we take to be equal to the top quark mass, $m_t$.

The search for and the setting of limits on anomalous top-quark couplings is an active area of LHC physics \cite{CMS,ATLAS}. In the determination of experimental limits it is important to have good theoretical predictions. When higher-order corrections are large, their effect is important in the setting of limits on anomalous couplings.

As in any process, higher-order radiative corrections may significantly change 
the leading-order (LO) result. The next-to-leading order (NLO) corrections to $tZ$ production via anomalous couplings were calculated in Ref. \cite{LZ} and were shown to be quite large. Therefore it is important to see if next-to-next-to-leading order (NNLO) corrections have significant effects on the production cross section.

A particular class of radiative corrections that have been well known to be significant and in fact dominant for massive final states with top quarks is from soft-gluon emission \cite{NKBP,NKHQ13,NKtop,NKts,NKtW,NKAB,NKGS,NKnnlo}. For example, in $tW$ production, which is a similar process, the soft-gluon corrections approximate very well the exact results at NLO. Denoting the LO result plus the soft corrections at NLO as approximate NLO (aNLO), it was found that the difference between the full NLO and the aNLO cross sections is negligible. The NNLO soft-gluon corrections contribute significantly to the cross section and the top-quark differential distributions \cite{NKtW}. The sum of the NLO result plus the NNLO soft corrections is denoted as approximate NNLO (aNNLO). It is reasonable to expect a similar behavior for $tZ$ production, i.e. a considerably enhanced aNNLO total cross section and differential distributions.

The soft-gluon corrections are particularly significant near partonic threshold, where by definition there is little energy left for any additional radiation and hence any gluon emission is soft, i.e. low energy. Soft-gluon radiation contributions appear in the form of logarithmic ``plus'' distributions of the type
\beq
\left[\frac{\ln^{k}(s_4/m_t^2)}{s_4} \right]_+ \, , 
\label{lnplus}
\eeq
with $k\le 2n-1$ at $n$th order in the strong coupling $\alpha_s$. 
These plus distributions arise from cancellations of 
infrared divergences between soft and virtual terms.
Double logarithms are from collinear and soft emission, 
while purely soft emission contributes single logarithms.
As we will see, the soft-gluon corrections for $tZ$ production are indeed large. We also note that while there may be different motivations and models for top-quark anomalous couplings, the effect of soft-gluon corrections is not dependent on the details of such models.

In the next section we present the soft-gluon resummation formalism and details of its application to $tZ$ production. Expansions of the resummed cross section are derived at NLO and NNLO. In Section 3 we present numerical results for the total $tZ$ production cross section as well as the top-quark transverse momentum, $p_T$, and rapidity distributions through aNNLO at LHC energies. We conclude in Section 4.

\begin{figure}
\begin{center}
\includegraphics[width=11cm]{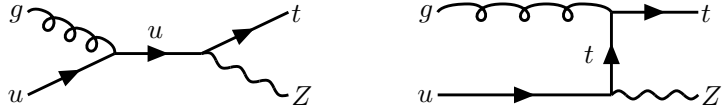}
\caption{Leading-order diagrams for the process $gu \rightarrow tZ$ via anomalous couplings.}
\label{lo-tz}
\end{center}
\end{figure}

\mysection{Resummation for $gq \rightarrow t Z$}

We study soft-gluon resummation for $tZ$ production via anomalous $t$-$q$-$Z$
couplings. The partonic process is $gq \rightarrow t Z$ where $q$ can be an
up or charm quark. The lowest-order Feynman diagrams with an up quark are
shown in Fig. \ref{lo-tz}; the same diagrams apply to the charm-quark
initiated process.

For the partonic process $g(p_g)+q(p_q) \rightarrow t(p_t)+Z(p_Z)$,
we define the kinematical variables $s=(p_g+p_q)^2$,
$t=(p_g-p_t)^2$, $u=(p_q-p_t)^2$, and $s_4=s+t+u-m_t^2-m_Z^2$,
where $m_Z$ is the $Z$-boson mass and, as before, $m_t$ is the top-quark mass. 
Near partonic threshold, i.e. when there is just enough
energy to produce the final $tZ$ state - but with the top-quark and $Z$-boson
not necessarily at rest,  we have $s_4 \rightarrow 0$.
We also define $t_1=t-m_t^2$, $t_2=t-m_Z^2$, $u_1=u-m_t^2$, and $u_2=u-m_Z^2$.

We consider the double-differential partonic cross section
$d^2{\hat\sigma^{(n)}}_{gq \rightarrow t Z}/(dt \, du)$ at $n$th order.
The LO cross section is 
\beq
\frac{d^2{\hat\sigma^{(0)}}_{gq \rightarrow t Z}}{dt \, du}
=F^{\rm LO}_{gq \rightarrow t Z} \, \delta(s_4) \, ,
\label{LO}
\eeq
where 
\beqa
F^{\rm LO}_{gq \rightarrow t Z}&=&
\frac{2 \pi \alpha \alpha_s \kappa_{tqZ}^2}{3s^3 t_1^2}
\left\{2 m_t^6 -m_t^4(3 m_Z^2+4 s+2 t) \right. 
\nonumber \\ &&
{}+m_t^2\left[2m_Z^4-m_Z^2(2s+t)+2(s^2+4st+t^2)\right] 
\nonumber \\ && 
{}+2m_Z^6-4m_Z^4t+m_Z^2(s+t)(s+5t)-2t(3s^2+6st+t^2)
\nonumber \\ && \left. 
{}-\frac{t}{m_t^2}\left[2 m_Z^6-2m_Z^4(s+t)
+m_Z^2(s+t)^2-4 s t(s+t)\right]\right\} \, ,
\eeqa
with $\alpha=e^2/(4\pi)$.

Resummation of soft-gluon corrections follows from the factorization of the cross section into functions that describe soft and collinear emission in the partonic process. Taking moments of the partonic cross section,  
${\hat \sigma}(N)=\int (ds_4/s) \;  e^{-N s_4/s} {\hat \sigma}(s_4)$, with $N$ the moment variable, we write a factorized expression in $4-\epsilon$ dimensions,
\beq
\frac{d^2{\hat \sigma}_{gq \rightarrow tZ}(N,\epsilon)}{dt \, du}= 
H_{gq \rightarrow tZ} \left(\alpha_s(\mu)\right)\; S_{gq \rightarrow tZ} 
\left(\frac{m_t}{N \mu},\alpha_s(\mu) \right)\;
\prod_{i=g,q} J_i\left (N,\mu,\epsilon \right) 
\label{factsigma}
\eeq 
where $\mu$ is the scale, $H_{gq \rightarrow tZ}$ is the hard-scattering function, 
$S_{gq \rightarrow tZ}$ is the soft-gluon function that encompasses 
noncollinear soft-gluon emission, 
and $J_i$ are jet functions which describe 
soft and collinear emission from the incoming quark and gluon.

The soft function $S_{gq \rightarrow tZ}$ requires renormalization;
its dependence on the moment $N$ is resummed via renormalization
group evolution \cite{NKtW,NKGS}. We write
\beq
S^b_{gq \rightarrow tZ}=(Z^S)^* \; S_{gq \rightarrow tZ} \, Z^S
\eeq
with $S^b_{gq \rightarrow tZ}$ the unrenormalized bare quantity, and $Z^S$ a renormalization constant. The function $S_{gq \rightarrow tZ}$ satisfies the renormalization group equation
\beq
\left(\mu \frac{\partial}{\partial \mu}
+\beta(g_s, \epsilon)\frac{\partial}{\partial g_s}\right)\,S_{gq \rightarrow tZ}
=-2 \, S_{gq \rightarrow tZ} \, \Gamma^S_{gq \rightarrow tZ}
\eeq
where $g_s^2=4\pi\alpha_s$; 
$\beta(g_s, \epsilon)=-g_s \epsilon/2 + \beta(g_s)$ 
with $\beta(g_s)$ the QCD beta function; and 
\beq
\Gamma^S_{gq \rightarrow tZ}=\frac{dZ^S}{d\ln\mu} (Z^S)^{-1}
=\beta(g_s, \epsilon)  \frac{\partial Z^S}{\partial g_s} (Z^S)^{-1}
\eeq
is the soft anomalous dimension that determines the evolution of the 
soft function $S_{gq \rightarrow tZ}$.
We calculate the soft anomalous dimension $\Gamma^S_{gq \rightarrow tZ}$ from the coefficients of the ultraviolet poles of the corresponding loop diagrams in dimensional regularization \cite{NKtop,NKts,NKtW,NKAB,NKGS,NK2loop}.

The resummed partonic cross section in moment space 
is then given by 
\beqa
\frac{d^2{\hat{\sigma}}^{\rm resum}_{gq \rightarrow tZ}(N)}{dt \, du} &=&   
\exp\left[\sum_{i=g,q} E_i(N_i)\right]
H_{gq \rightarrow tZ}
\left(\alpha_s(\sqrt{s})\right) \;
S_{gq \rightarrow tZ}\left(\alpha_s(\sqrt{s}/{\tilde N'})
\right) 
\nonumber \\ && 
\times \exp \left[2\int_{\sqrt{s}}^{{\sqrt{s}}/{\tilde N'}} 
\frac{d\mu}{\mu}\; \Gamma^S_{gq \rightarrow tZ}
\left(\alpha_s(\mu)\right)\right]  \, .
\label{resum}
\eeqa
The first exponent in Eq. (\ref{resum}) resums soft and collinear 
corrections \cite{GS87,CT89} from the incoming quark and gluon,
and it can be found explicitly in \cite{NKtW}.

We expand the soft anomalous dimension for $gq \rightarrow tZ$ as
$\Gamma^S_{gq \rightarrow tZ}=\sum_{n=1}^{\infty}(\alpha_s/\pi)^n
\Gamma^{S \, (n)}_{gq \rightarrow tZ}$.
For resummation at next-to-leading-logarithm (NLL) accuracy we need the 
expression for the soft anomalous dimension at one loop. 
The one-loop result in Feynman gauge is 
\beq
\Gamma^{S\, (1)}_{gq \rightarrow tZ}=
C_F \left[\ln\left(\frac{-u_1}{m_t\sqrt{s}}\right)
-\frac{1}{2}\right] +\frac{C_A}{2} \ln\left(\frac{t_1}{u_1}\right)
\label{tZ1l}
\eeq
with color factors $C_F=(N_c^2-1)/(2N_c)$ and $C_A=N_c$, 
where $N_c=3$ is the number of colors.

We have also calculated the two-loop result, which is 
\beq
\Gamma^{S\, (2)}_{gq \rightarrow tZ}=
\left[C_A\left(\frac{67}{36}-\frac{\zeta_2}{2}\right)
  -\frac{5}{18}n_f\right]  \Gamma^{S\, (1)}_{gq \rightarrow tZ}
+C_F C_A \frac{(1-\zeta_3)}{4}
\label{tZ2l}
\eeq
where $\Gamma^{S\, (1)}_{gq \rightarrow tZ}$ is given in Eq. (\ref{tZ1l}),  
$n_f=5$ is the number of light-quark flavors,
$\zeta_2=\pi^2/6$, and $\zeta_3=1.2020569\cdots$.

The expansion of the NLL resummed cross section, Eq. (\ref{resum}), to NNLO \cite{NKnnlo} followed by inversion to momentum space, provides us with robust and prescription-independent results for the soft-gluon corrections at NLO and NNLO. As we will see in the next section, the NLO soft-gluon corrections are large and they approximate the complete corrections at that order very well, while the NNLO soft-gluon corrections provide further significant contributions. 

The NLO soft-gluon corrections for $g q \rightarrow tZ$ are
\beqa
\frac{d^2{\hat\sigma}^{(1)}_{gq\rightarrow t Z}}{dt \, du}
&=&F^{\rm LO}_{gq \rightarrow t Z} 
\frac{\alpha_s(\mu_R^2)}{\pi} \left\{
2 (C_F+C_A) \left[\frac{\ln(s_4/m_t^2)}{s_4}\right]_+ \right.
\nonumber \\ && \hspace{-23mm}
{}+\left[2 C_F \ln\left(\frac{u_1}{t_2}\right)-C_F
+C_A \ln\left(\frac{t_1}{u_1}\right)
+C_A \ln\left(\frac{s m_t^2}{u_2^2}\right)
-(C_F+C_A)\ln\left(\frac{\mu_F^2}{m_t^2}\right)\right] 
\left[\frac{1}{s_4}\right]_+
\nonumber \\ && \hspace{-23mm} \left.
{}+\left[\left(C_F \ln\left(\frac{-t_2}{m_t^2}\right)
+C_A \ln\left(\frac{-u_2}{m_t^2}\right)
-\frac{3}{4}C_F\right)\ln\left(\frac{\mu_F^2}{m_t^2}\right)
-\frac{\beta_0}{4}\ln\left(\frac{\mu_F^2}{\mu_R^2}\right)\right]  
\delta(s_4)\right\} \, ,
\label{NLOgqtZ}
\eeqa
where $\mu_F$ is the factorization scale, $\mu_R$ is the renormalization scale 
and $\beta_0=(11C_A-2n_f)/3$ is the lowest-order QCD $\beta$ function.

The NNLO soft-gluon corrections for $g q \rightarrow tZ$ are
\beqa
\frac{d^2{\hat\sigma}^{(2)}_{gq\rightarrow t Z}}{dt \, du}
&=&F^{\rm LO}_{gq \rightarrow t Z} 
\frac{\alpha_s^2(\mu_R^2)}{\pi^2} 2 (C_F+C_A) \left\{
(C_F+C_A) \left[\frac{\ln^3(s_4/m_t^2)}{s_4}\right]_+ \right.
\nonumber \\ && \hspace{-25mm}
{}+\frac{3}{2}\left[2 C_F \ln\left(\frac{u_1}{t_2}\right)-C_F
+C_A \ln\left(\frac{t_1}{u_1}\right)
+C_A \ln\left(\frac{s m_t^2}{u_2^2}\right)
-(C_F+C_A)\ln\left(\frac{\mu_F^2}{m_t^2}\right) -\frac{\beta_0}{6}\right]
\left[\frac{\ln^2(s_4/m_t^2)}{s_4}\right]_+
\nonumber \\ && \hspace{-25mm}
{}+\left[\left(3C_F \ln\left(\frac{-t_2}{m_t^2}\right)
-2C_F \ln\left(\frac{-u_1}{m_t^2}\right)+\frac{C_F}{4}
+3C_A \ln\left(\frac{-u_2}{m_t^2}\right)
+C_A \ln\left(\frac{u_1 m_t^2}{t_1 s}\right)-\frac{\beta_0}{4}\right)
\ln\left(\frac{\mu_F^2}{m_t^2}\right) \right.
\nonumber \\ && \hspace{-18mm} \left. 
{}+\frac{\beta_0}{2} \ln\left(\frac{\mu_R^2}{m_t^2}\right)
+\frac{1}{2}(C_F+C_A) \ln^2\left(\frac{\mu_F^2}{m_t^2}\right)\right]
\left[\frac{\ln(s_4/m_t^2)}{s_4}\right]_+
\nonumber \\ && \hspace{-25mm} \left.
{}+\left[-\frac{\beta_0}{4}\ln\left(\frac{\mu_F^2}{m_t^2}\right)\ln\left(\frac{\mu_R^2}{m_t^2}\right) 
+\left(\frac{3\beta_0}{16}+\frac{3}{8}C_F-\frac{C_F}{2}\ln\left(\frac{-t_2}{m_t^2}\right)
-\frac{C_A}{2}\ln\left(\frac{-u_2}{m_t^2}\right)\right)\ln^2\left(\frac{\mu_F^2}{m_t^2}\right) \right] \left[\frac{1}{s_4}\right]_+ \right\} \, .
\nonumber \\ 
\label{NNLOgqtZ}
\eeqa

The plus distributions that appear in the above corrections are defined by their integral with parton distributions $\phi$, as 
\beqa
\int_0^{s_{4 \, max}} ds_4 \, \phi(s_4) \left[\frac{\ln^k(s_4/m_t^2)}
{s_4}\right]_{+} &=&
\int_0^{s_{4\, max}} ds_4 \frac{\ln^k(s_4/m_t^2)}{s_4} [\phi(s_4) - \phi(0)]
\nonumber \\ &&
{}+\frac{1}{k+1} \ln^{k+1}\left(\frac{s_{4\, max}}{m_t^2}\right) \phi(0) \, .
\label{splus}
\eeqa

\begin{figure}
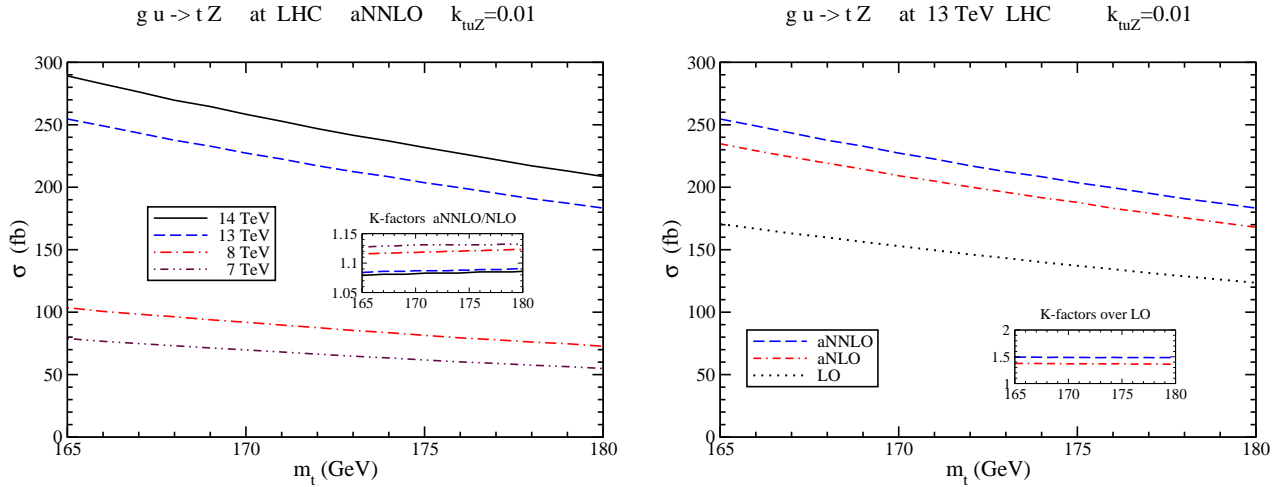

\begin{center}
\includegraphics[width=81mm]{gutZlhcv2plot.eps}
\hspace{3mm}
\includegraphics[width=81mm]{gutZ13lhcv2plot.eps}
\caption{Total cross sections for $tZ$ production via the process $gu\rightarrow tZ$ with anomalous $t$-$u$-$Z$ coupling at LHC energies. The plot on the left shows aNNLO results at 7, 8, 13, and 14 TeV. The plot on the right shows LO, aNLO, and aNNLO results at 13 TeV energy. The inset plots display $K$-factors as discussed in the text.}
\label{gutZ}
\end{center}
\end{figure}

\mysection{Results for $gu\rightarrow tZ$ and $gc\rightarrow tZ$ at the LHC}

We now present numerical results for $tZ$ production in $pp$ collisions at LHC energies via top-quark $t$-$q$-$Z$ anomalous couplings, specifically via the partonic processes $gu\rightarrow tZ$ and $gc\rightarrow tZ$. Throughout we use CT14 \cite{CT14} NNLO parton distribution functions (pdf), but we note that the results are essentially the same if MMHT2014 \cite{MMHT} or NNPDF \cite{NNPDF} pdf are used instead.

We first note that we have compared our aNLO results with the complete NLO results of Ref. \cite{LZ}. We find remarkable agreement, with only a few per mille difference between the aNLO and NLO cross sections for $gu \rightarrow tZ$, and less than two percent difference for $gc \rightarrow tZ$. This shows that the soft-gluon corrections dominate the corrections and thus serve as an excellent approximation to the complete corrections. This is in line with studies for other top-quark processes, and in particular for $tW$ production \cite{NKtW}, which has a similar color structure and mass for the final-state particles. In addition, our aNLO results for the top $p_T$ distributions are very close to the exact NLO results of Ref. \cite{LZ}, at the few percent or better level for both $gu \rightarrow tZ$ and $gc \rightarrow tZ$, again in line with expectations. Given these considerations, it is expected that our aNNLO corrections should provide a reliable measure of the contributions beyond NLO.

We begin with the process $gu\rightarrow tZ$. For specificity, we choose a value for the anomalous coupling of $\kappa_{tuZ}=0.01$, in line with recent limits \cite{CMS,ATLAS}. In the left plot of Fig. \ref{gutZ} we show the aNNLO total cross sections for $gu\rightarrow tZ$ in $pp$ collisions at 7, 8, 13, and 14 TeV LHC energies as functions of top-quark mass in the range from 165 to 180 GeV. The factorization and renormalization scales are set equal to the top-quark mass.

The inset in the left plot of Fig. \ref{gutZ} shows the $K$-factors, i.e. the aNNLO/NLO cross section ratios at the various LHC energies. We see that the aNNLO corrections increase the NLO result by around 8\% at 14 TeV, 9\% at 13 TeV, 12\% at 8 TeV, and 13\% at 7 TeV. The increasing effect of the corrections as the energy is lowered is entirely expected as we get closer to partonic threshold, where the contribution of the soft-gluon corrections gets larger. 

The plot on the right of Fig. \ref{gutZ} shows the soft-gluon contributions at each order to the total cross section for $gu\rightarrow tZ$ at 13 TeV energy. The LO, aNLO, and aNNLO cross sections are displayed, all with the same pdf, thus showing the effect of the perturbative soft-gluon corrections. It is evident that the NLO soft-gluon corrections substantially increase the LO result and, as we discussed above, dominate the complete NLO cross section. The aNNLO soft-gluon corrections further increase the cross section and need to be included for a more precise theoretical prediction.

The inset in the right plot of Fig. \ref{gutZ} shows the $K$-factors relative to LO, i.e. the ratios of the higher-order cross sections to the LO cross section. The aNLO/LO $K$-factor at 13 TeV is 1.37, indicating 37\% enhancement of the LO cross section by the NLO soft-gluon corrections. The aNNLO/LO $K$-factor is 1.49, indicating an additional 12\% enhancement from the NNLO soft-gluon corrections. 

\begin{figure}
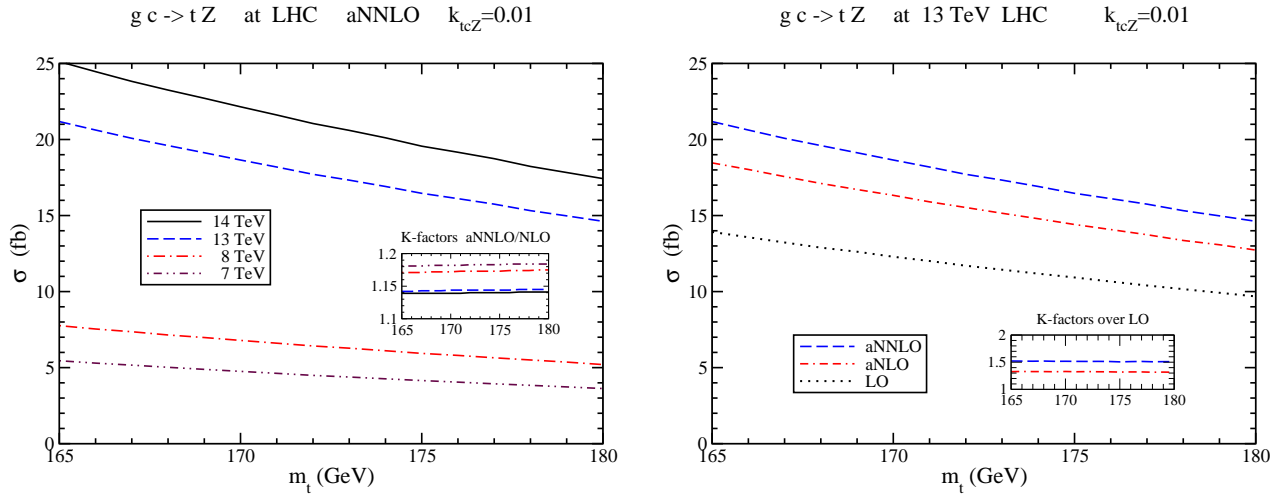

\begin{center}
\includegraphics[width=81mm]{gctZlhcv2plot.eps}
\hspace{3mm}
\includegraphics[width=81mm]{gctZ13lhcv2plot.eps}
\caption{Total cross sections for $tZ$ production via the process $gc\rightarrow tZ$ with anomalous $t$-$c$-$Z$ coupling at LHC energies. The plot on the left shows aNNLO results at 7, 8, 13, and 14 TeV. The plot on the right shows LO, aNLO, and aNNLO results at 13 TeV energy. The inset plots display $K$-factors as discussed in the text.}
\label{gctZ}
\end{center}
\end{figure}

We continue with the process $gc\rightarrow tZ$. We choose a value for the anomalous coupling of $\kappa_{tcZ}=0.01$. The cross section for this process is an order of magnitude smaller than the one with up quarks. In the left plot of Fig. \ref{gctZ} we show the aNNLO total cross sections for $gc\rightarrow tZ$ in $pp$ collisions at 7, 8, 13, and 14 TeV LHC energies as functions of top-quark mass, again in the range from 165 to 180 GeV. The inset plot on the left shows the aNNLO/NLO $K$-factors at different energies. The $K$-factors are again large, in fact even larger than for the up-quark initiated process. Again, we observe that the $K$-factors get larger at lower energies, as expected.

The plot on the right of Fig. \ref{gctZ} shows the LO, aNLO, and aNNLO cross sections for $gc\rightarrow tZ$ at 13 TeV energy. We again observe that the NLO soft-gluon corrections substantially increase the LO result and that the aNNLO soft-gluon corrections further increase the cross section: the aNNLO/LO $K$-factor is 1.51. These are very significant corrections, similar to what we found for the process $gu\rightarrow tZ$.

Since the higher-order corrections are large, the effect on setting limits on anomalous couplings is significant. For the $t$-$u$-$Z$ coupling at 13 TeV, the NLO calculation reduces the limit on the coupling by 17\% relative to LO, while the aNNLO calculation reduces the limit by 22\% relative to LO. For the $t$-$c$-$Z$ coupling at 13 TeV, the NLO calculation reduces the limit on the coupling by 15\% relative to LO, while the aNNLO calculation reduces the limit by 23\% relative to LO.

\begin{figure}
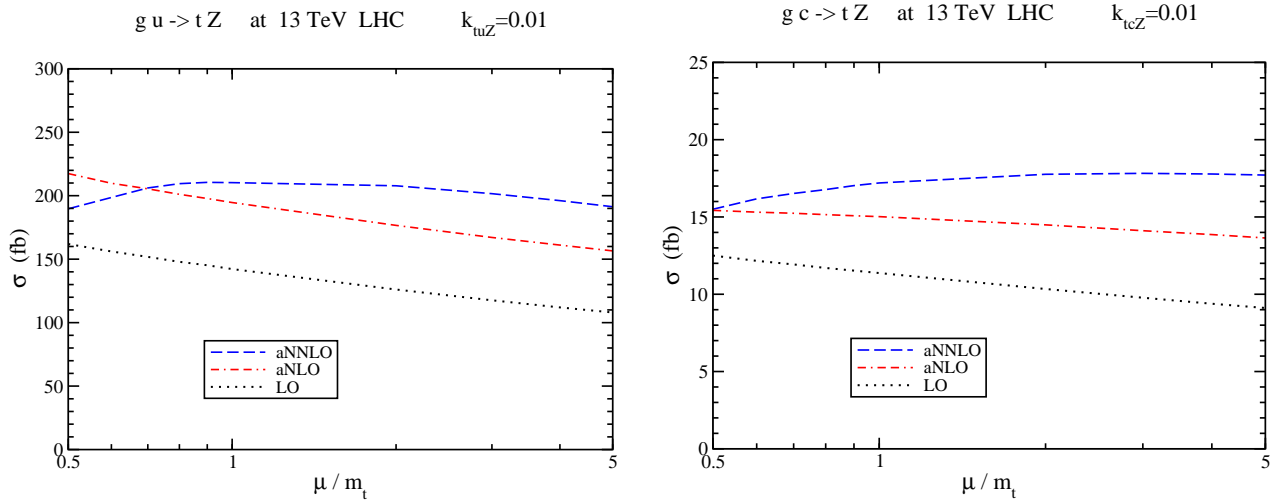

\begin{center}
\includegraphics[width=81mm]{gutZ13lhcmuv2plot.eps}
\hspace{3mm}
\includegraphics[width=81mm]{gctZ13lhcmuv2plot.eps}
\caption{The scale dependence of the total cross sections for $tZ$ production via the processes $gu\rightarrow tZ$ (left) and $gc\rightarrow tZ$ (right) at 13 TeV LHC energy with $m_t=173.3$ GeV.}
\label{gutZmu}
\end{center}
\end{figure}

In Figure \ref{gutZmu} we show the scale dependence of the cross sections for the current 13 TeV energy at the LHC. The top-quark mass is taken to be 173.3 GeV. The left plot is for the process $gu \rightarrow tZ$, while the right plot is for $gc \rightarrow tZ$. We observe a relatively mild dependence of the cross section on scale, especially when the aNNLO corrections are included. Typically, a scale variation by a factor of two is used to show an estimate of the theoretical uncertainty, but we show a bigger range in the plot. 

For the process $gu \rightarrow tZ$ at 13 TeV energy, there is a scale uncertainty of 5.2\% at aNNLO in the range shown, which is much smaller than the 16.3\% uncertainty at NLO and the 20.0\% uncertainty at LO. Thus, the inclusion of the higher-order corrections greatly reduces the theoretical uncertainty. The corresponding aNNLO uncertainties for other LHC energies are similar: 5.9\% at 14 TeV, 6.3\% at 8 TeV, and 6.6\% at 7 TeV. 

For the process $gc \rightarrow tZ$ at 13 TeV energy, there is a scale uncertainty of 7.0\% at aNNLO in the range shown, which is much smaller than the 15.7\% uncertainty at LO. We also find an aNNLO scale uncertainty of 6.8\% at 14 TeV, 4.7\% at 8 TeV, and 4.2\% at 7 TeV.

In addition to scale uncertainties, there are also pdf uncertainties. Using the pdf errors associated with the CT14NNLO pdf \cite{CT14}, we find that the pdf uncertainties are relatively small for $gu \rightarrow tZ$: they are 3.0\% at 14 TeV, 2.9\% at 13 TeV, and 2.8\% at 7 and 8 TeV. The pdf uncertainties are larger for the process $gc \rightarrow tZ$: 6.0\% at 14 TeV, 6.3\% at 13 TeV, 8.9\% at 8 TeV, and 9.8\% at 7 TeV.

There is also the question of uncertainties from missing terms beyond the ones already included. The full set of those missing terms are of course not known, but one way to check their possible impact is to include certain terms beyond NLL accuracy that are known. In particular, a theoretically important term is the two-loop soft anomalous dimension, Eq. (\ref{tZ2l}). Including it in the aNNLO expression, we find that numerically its effect is less than one per mille for the total cross section, which is tiny compared to the scale and pdf uncertainties.

Next we consider the top-quark differential distributions. In particular, we calculate the top-quark transverse-momentum ($p_T$) and rapidity distributions at LHC energies through aNNLO. We set the scales equal to the top-quark mass which we again take as 173.3 GeV.

\begin{figure}
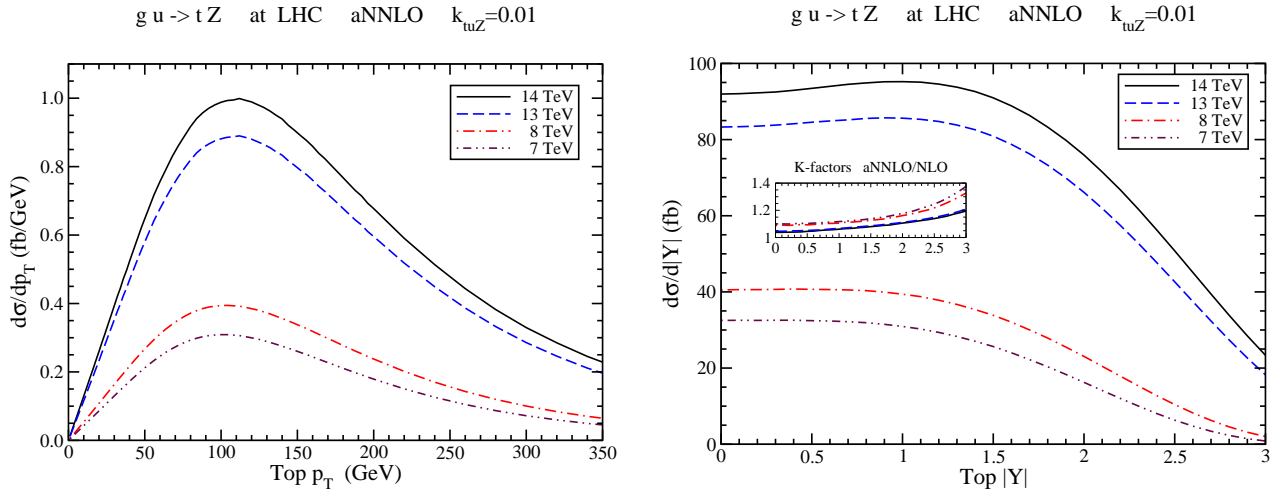

\begin{center}
\includegraphics[width=81mm]{pttopgutZlhcv2plot.eps}
\hspace{3mm}
\includegraphics[width=81mm]{yabstopgutZlhcv2plot.eps}
\caption{The top-quark $p_T$ (left) and rapidity (right) distributions in the process $gu\rightarrow tZ$ at LHC energies with $m_t=173.3$ GeV.}
\label{ptygutZ}
\end{center}
\end{figure}

We begin with the process $gu\rightarrow tZ$. In the left plot of Fig. \ref{ptygutZ} we show the aNNLO top-quark $p_T$ distributions, $d\sigma/dp_T$, at 7, 8, 13, and 14 TeV LHC energies with $m_t=173.3$ GeV. The $p_T$ distributions peak at a $p_T$ value of around 110 GeV for 13 and 14 TeV energy, and at around 100 GeV for 7 and 8 TeV energy. The corresponding top-quark rapidity distributions, $d\sigma/d|Y|$, are shown in the plot on the right of the figure. We note that the soft-gluon corrections for the rapidity distribution are increasingly significant at large rapidity values, as is expected at edges of kinematic phase space. As the $K$-factors in the inset of the right plot show, the aNNLO corrections can be very large, increasing the NLO rapidity distribution at a rapidity value $|Y|=3$ by around 20\% at 14 TeV, 21\% at 13 TeV, 33\% at 8 TeV, and 38\% at 7 TeV. As also expected, the $K$-factors increase with decreasing energy. 

\begin{figure}
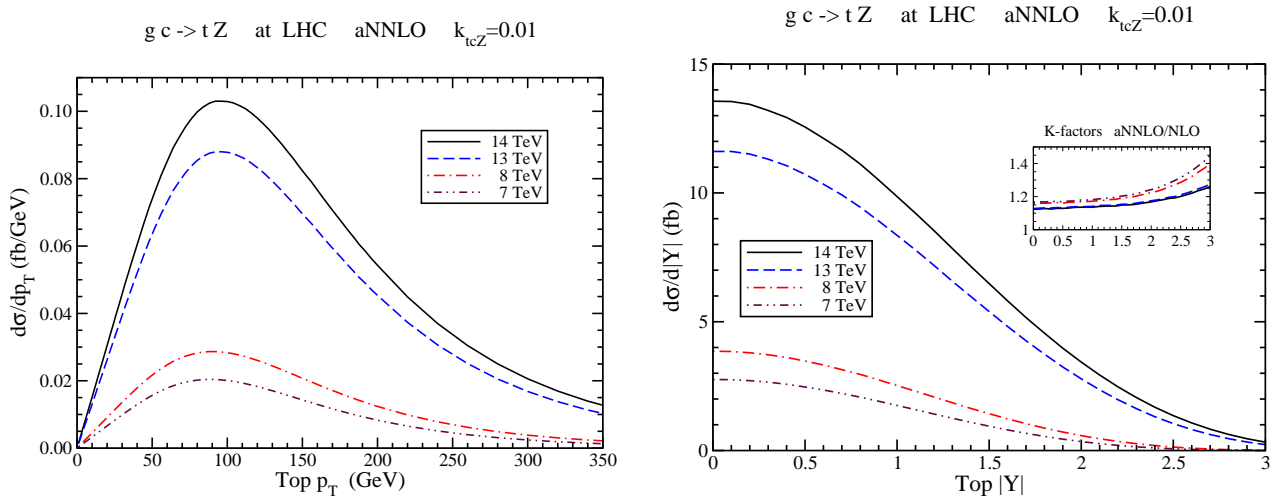

\begin{center}
\includegraphics[width=81mm]{pttopgctZlhcv2plot.eps}
\hspace{3mm}
\includegraphics[width=81mm]{yabstopgctZlhcv2plot.eps}
\caption{The top-quark $p_T$ (left) and rapidity (right) distributions in the process $gc\rightarrow tZ$ at LHC energies with $m_t=173.3$ GeV.}
\label{ptygctZ}
\end{center}
\end{figure}

Finally, we present the top-quark differential distributions in the process $gc\rightarrow tZ$, again with $m_t=173.3$ GeV. As for the total cross sections, the differential distributions are an order of magnitude smaller than for the up-quark initiated process. In the left plot of Fig. \ref{ptygctZ} we show the aNNLO top-quark $p_T$ distributions, and on the right plot we show the rapidity distributions at LHC energies. The $K$-factors in the inset of the right plot again show a similar pattern, increasing both at higher rapidities and lower energies. The aNNLO corrections increase the NLO rapidity distribution at a rapidity value $|Y|=3$ by around 26\% at 14 TeV, 28\% at 13 TeV, 40\% at 8 TeV, and 45\% at 7 TeV.

\mysection{Conclusions}

The production of a top quark in association with a $Z$ boson can proceed via anomalous $t$-$q$-$Z$ top-quark couplings, where $q$ is an up or charm quark. The partonic processes involved are $gu \rightarrow tZ$ and $gc \rightarrow tZ$. We have shown that the total cross sections as well as the top-quark transverse momentum and rapidity distributions receive large contributions from soft-gluon emission; these contributions dominate the cross section numerically. We resummed soft and collinear gluon emission via the use of soft anomalous dimensions which have been calculated to two loops. From the resummed cross section at NLL accuracy we derived expansions at NLO and NNLO. The soft-gluon corrections at NLO approximate very well the exact results. The aNNLO soft-gluon corrections provide substantial additional enhancements.

The total production cross sections for both the $gu \rightarrow tZ$ and $gc \rightarrow tZ$ partonic processes were calculated at 7, 8, 13, and 14 TeV LHC energies. The aNNLO/NLO $K$-factors show that the aNNLO corrections are significant at all energies. These contributions need to be included in searching for anomalous couplings and in setting limits on them at the LHC. Furthermore, the top-quark transverse momentum and rapidity distributions were calculated in both processes. The aNNLO corrections are particularly large for large values of the top-quark rapidity.

\mysection*{Acknowledgements}
This material is based upon work supported by the National Science Foundation under Grant No. PHY 1519606.


\begin{thebibliography}{99}

\bibitem{WB}
W. Bernreuther, J. Phys. G {\bf 35}, 083001 (2008) [arXiv:0805.1333 [hep-ph]].

\bibitem{DW}
D. Wackeroth, in {\sl ICHEP 2008}, arXiv:0810.4176 [hep-ph].

\bibitem{IQWW}
J.R. Incandela, A. Quadt, W. Wagner, and D. Wicke, 
Prog. Part. Nucl. Phys. {\bf 63}, 239 (2009)
[arXiv:0904.2499 [hep-ex]].

\bibitem{AHTJ}
A. Heinson and T.R. Junk, 
Ann. Rev. Nucl. Part. Sci. {\bf 61}, 171 (2011) 
[arXiv:1101.1275 [hep-ex]]. 

\bibitem{NKBP}
N. Kidonakis and B.D. Pecjak,  
Eur. Phys. J. C {\bf 72}, 2084 (2012) [arXiv:1108.6063 [hep-ph]].

\bibitem{RBAF}
R. Bonciani and A. Ferroglia, PoS (EPS-HEP2011) 341 [arXiv:1201.4382 [hep-ph]].

\bibitem{PF}
P. Falgari, J. Phys. Conf. Ser. {\bf 452}, 012016 (2013)
[arXiv:1302.3699 [hep-ph]]. 

\bibitem{NKHQ13}
N. Kidonakis, in Proceedings of the Helmholtz International School Physics of Heavy Quarks and Hadrons, HQ2013, Dubna, Russia, 2013, DESY-PROC-2013-03, p. 139 [arXiv:1311.0283 [hep-ph]]. 

\bibitem{DDL}
V. Del Duca and E. Laenen, Int. J. Mod. Phys. A {\bf 30}, 1530063 (2015) [arXiv:1510.06690 [hep-ph]].

\bibitem{AG}
A. Giammanco, Rev. Phys. {\bf 1}, 1 (2016) [arXiv:1511.06748 [hep-ex]].

\bibitem{JWK}
J. Wagner-Kuhr, arXiv:1606.02936 [hep-ex].

\bibitem{AGRS}
A. Giammanco and R. Schwienhorst, arXiv:1710.10699 [hep-ex].

\bibitem{NKtop}	
N. Kidonakis, Phys. Rev. D {\bf 90}, 014006 (2014) [arXiv:1405.7046 [hep-ph]];
Phys. Rev. D {\bf 91}, 031501(R) (2015) [arXiv:1411.2633 [hep-ph]]; 
Phys. Rev. D {\bf 91}, 071502(R) (2015) [arXiv:1501.01581 [hep-ph]].

\bibitem{NKts}
N. Kidonakis, Phys. Rev. D {\bf 81}, 054028 (2010) [arXiv:1001.5034 [hep-ph];
Phys. Rev. D {\bf 83}, 091503(R) (2011) [arXiv:1103.2792 [hep-ph]]; 
Phys. Rev. D {\bf 93}, 054022 (2016) [arXiv:1510.06361 [hep-ph]].

\bibitem{NKtW}
N. Kidonakis, Phys. Rev. D {\bf 82}, 054018 (2010) [arXiv:1005.4451 [hep-ph]];
Phys. Rev. D {\bf 96}, 034014 (2017) [arXiv:1612.06426 [hep-ph]].

\bibitem{TY}
T. Tait and C.-P. Yuan, Phys. Rev. D {\bf 63}, 014018 (2000) [hep-ph/0007298].

\bibitem{NKAB}
N. Kidonakis and A. Belyaev, JHEP {\bf 12}, 004 (2003) [hep-ph/0310299]. 

\bibitem{LZ}
B.H. Li, Y. Zhang, C.S. Li, J. Gao, and H.X. Zhu, Phys. Rev. D {\bf 83}, 114049 (2011) [arXiv:1103.5122 [hep-ph]]. 

\bibitem{AAC}
J.-L. Agram, J. Andrea, E. Conte, B. Fuks, D. Gele, and P. Lansonneur, Phys. Lett. B {\bf 725}, 123 (2013) [arXiv:1304.5551 [hep-ph]].

\bibitem{TopC}
J. Adelman {\it et al.}, arXiv:1309.1947 [hep-ex].

\bibitem{DMZ}
G. Durieux, F. Maltoni, and C. Zhang, Phys. Rev. D {\bf 91}, 074017 (2015) [arXiv:1412.7166 [hep-ph]]. 

\bibitem{SMtZ}
J. Campbell, R.K. Ellis, and R. Rontsch, Phys. Rev. D {\bf 87} (2013) 114006 (2013) [arXiv:1302.3856 [hep-ph]]. 

\bibitem{CMS}
CMS Collaboration, JHEP {\bf 07}, 003 (2017) [arXiv:1702.01404 [hep-ex]].

\bibitem{ATLAS}
ATLAS Collaboration, ATLAS-CONF-2017-070.

\bibitem{NKGS}
N. Kidonakis and G. Sterman,  Nucl. Phys. B {\bf 505}, 321 (1997) [hep-ph/9705234].

\bibitem{NKnnlo}
N. Kidonakis, Mod. Phys. Lett. A {\bf 19}, 405 (2004) [hep-ph/0401147]. 

\bibitem{NK2loop}
N. Kidonakis, Phys. Rev. Lett. {\bf 102}, 232003 (2009) [arXiv:0903.2561 [hep-ph]]. 

\bibitem{GS87}
G. Sterman, Nucl. Phys. B {\bf 281}, 310 (1987). 

\bibitem{CT89}
S. Catani and L. Trentadue, Nucl. Phys. B {\bf 327}, 323 (1989).

\bibitem{CT14}
S. Dulat, T.-J. Hou, J. Gao, M. Guzzi, J. Huston, P. Nadolsky, J. Pumplin, 
C. Schmidt, D. Stump, and C.-P. Yuan, Phys. Rev. D {\bf 93}, 033006 (2016) 
[arXiv:1506.07443 [hep-ph]].

\bibitem{MMHT}
L.A. Harland-Lang, A.D. Martin, P. Molytinski, and R.S. Thorne,   
Eur. Phys. J. C {\bf 75}, 204 (2015) [arXiv:1412.3989 [hep-ph]].

\bibitem{NNPDF}
NNPDF Collaboration, R.D. Ball {\it et al.}, 
Eur. Phys. J. C {\bf 77}, 663 (2017) [arXiv:1706.00428 [hep-ph]]. 

\end{thebibliography}
\end{document}